\documentclass[review]{elsarticle}
\usepackage{amsmath}
\usepackage{mathrsfs}
\usepackage{graphicx}
\usepackage{color}
\usepackage{mathtools}

\usepackage{lineno,hyperref}
\begin{document}

\begin{frontmatter}

\title{Numerical optimization of a low emittance lattice cell}

%% Group authors per affiliation:
\author[mymainaddress,mysecondaryaddress]{Tong Zhang}

\author[mysecondaryaddress]{Xiaobiao Huang}
\ead{xiahuang@slac.stanford.edu} 

\address[mymainaddress]{University of Science and Technology of China, Hefei, 
Anhui Province, China, 230000}
\address[mysecondaryaddress]{SLAC National Accelerator Laboratory, 2575 Sand Hill Road, Menlo Park, CA 94025}

\begin{abstract}
In the lattice designs for the next generation storage ring light sources, longitudinal 
gradient bending magnets and anti-bending magnets have been adopted. 
A logical question raised by the trend of varying the longitudinal distribution of dipole 
strength is: what are the optimal distributions of the dipole and quadrupole fields  
in a lattice cell for the purpose of minimizing the natural emittance? 
We studied this problem by numerically optimizing 
the dipole and quadrupole distributions of the normalized cell optics 
using the particle swarm optimization algorithm. 
The results reveal the features of the longitudinal field variation of the optimized 
cell and show that when the quadrupole gradient is increased enough, the cell 
tends to split into two identical cells with similar features. 
\end{abstract}


\end{frontmatter}

%\linenumbers

\section{Introduction}
Beam emittance is a critical performance parameter for synchrotron light sources 
and damping rings. 
Lower emittance in damping rings leads to higher luminosity in the associated linear 
colliders. 
For synchrotron light sources, lower emittance leads to higher photon beam brightness. 
The beam emittance in 
a high energy electron synchrotron is given by the beam energy and the magnetic lattice 
as the electron beam reaches an equilibrium distribution quickly through radiation damping 
and quantum excitation in the bending magnets. 
Because the ring lattice typically consists of periodic cells, the cell structure determines 
the emittance. 

The natural emittance of a storage ring with periodic cells may be expressed as 
\begin{eqnarray}\label{eqEmitForm1}
\epsilon_n=\frac{\mathcal{F}}{12\sqrt{15}}C_q\gamma^2\frac{\theta_c^3}{J_x},
\end{eqnarray}
where $C_q=3.83\times10^{-13}$~m, $\gamma$ is the Lorentz energy factor, 
$\theta_c=2\pi/N_c$ is the bending angle per cell, $N_c$ is the total number of cells, 
$J_x$ is the horizontal damping partition, and $\mathcal{F}$ is a dimensionless form 
factor for the cell type. 

Double-bend-achromat (DBA) lattice cells~\cite{DBAGreen1976} 
are the building blocks of most third generation 
light sources, with a few exceptions that use triple-bend-achromat (TBA) 
cells~\cite{TBAMurphy1992}.
The conditions for minimal emittances of these cell types have been 
analyzed in Ref.~\cite{SYLeeEmittancePRE96}.
The MAX-IV storage ring started a new trend of lattice design practice that adopts the 
multi-bend-achromat (MBA) cells~\cite{ERIKSSON2008221,LeemannPRSTAB2009}.
By using high gradient quadrupole magnets, the MBA lattice allows to 
focus the dispersion function down between bending magnets in shorter distances. This  
allows placing more ``focused'' bending magnets in a ring, which reduces 
dispersion in bending magnets and in turn reduces the emittance. 
Essentially an MBA cell consists of several smaller cells. If one bending magnet is considered as
one basic cell, then the MBA lattice substantially increases the number of cells, 
which is very effective in reducing the natural emittance as shown in Eq.~\eqref{eqEmitForm1}. 

In an MBA cell, the centers of the middle bending magnets are typically the minima 
of both the horizontal beta function and the dispersion function.  
Each middle bending magnet and 
its flanking quadrupoles resemble the theoretic minimum emittance (TME) 
cell structure~\cite{LTengTME1985}. 
Traditionally bending magnets have uniform magnetic field in the longitudinal direction, 
in which case the minimal form factor for a TME cell is found to be 
$\mathcal{F}_\text{TME}=1$. % \frac1{12\sqrt{15}}$. 

The middle bending magnets are often 
combined function magnets which also serve as defocusing quadrupoles. This saves space, and 
also reduces the horizontal emittance by increasing the damping partition. 
Bending magnets with varying dipole fields can be used 
to further reduce the beam emittance. 
The dipole field may be re-distributed to put stronger bending in the region with lower dispersion 
invariant and weaker bending elsewhere in order to achieve more evenly distributed 
quantum excitations throughout the magnet, which could in turn leads to smaller emittance. 
There have been many studies 
on the topic~\cite{GuoJEPAC02, PapaphiPAC05, NAGAOKA2007292, WangCXPRSTAB2009, 
STREUN201598}. 
In a recent study, Ref.~\cite{STREUN201598}, numeric optimization of the dipole field distribution 
for minimum emittance was conducted. 
Another important recent development in lattice design is the use of negative bends~\cite{AntiBendPac89, STREUN2014148}.
The negative bends allow a reduction of dispersion function in the main bending magnets, which 
further pushes down the emittance. 

The use of combined function magnets, longitudinal gradient bending magnets, and anti-bends are 
essentially a re-distribution of the bending and focusing functions on the cell length in order 
to achieve better emittance performance. 
This has inspired us to take a more general approach to investigate the low emittance linear lattice 
design problem. Suppose the focusing and bending functions are freely variable, subject to certain 
constraints, what would be the functions that minimize the emittance? 
Specifically for the synchrotron light sources, the actual performance measure of concern is the photon beam 
brightness, which not only depends on the natural emittance, but also the beta functions at the insertion devices. 
Since the beta functions are also lattice parameters, the problem becomes maximizing the brightness directly 
with the focusing and bending functions.  
In both cases, we have an optimization problem that can be solved numerically. 

In this study we used the particle swarm optimization (PSO) algorithm to minimize the emittance or maximize 
brightness
of a periodic storage ring lattice in the general case with free distributed focusing functions 
and bending functions over the length of a cell. Our results showed that  
the best approach to reduce emittance is to split the ring into as many 
identical cells as allowed by the constraint of the 
maximum quadrupole gradient. Within a cell, the bending function varies along the longitudinal 
direction, including a small section with negative bending. 

The paper is organized as follows. 
We first mathematically formulate the emittance minimization problem in Section~\ref{refNormCell}. 
It is followed by a description of the numeric optimization setup in Section~\ref{secOptimSetup}. 
The optimization results and discussions are presented in Section~\ref{secResults}. 
The conclusion is given in Section~\ref{sec:Conc}. 
 
\section{Optimization of emittance and brightness for a normalized lattice cell \label{refNormCell}}
\subsection{Normalization of a lattice cell}
Linear lattice functions of a periodic cell, such as beta functions and dispersion functions, are 
determined by the periodic conditions and the distributions of the quadrupole and dipole fields over the 
cell. A lattice cell can be scaled in length while keeping its linear lattice features intact. 
In fact, we can normalize the lattice functions using the length of the cell, $L$, as follows~\cite{HuangIPAC17}.  

Define normalized quantities
\begin{equation} \label{eq:er3}
\hat{s} = \frac{s}{L},\,\, 
\hat{\beta}_{x,y} = \frac{\beta_{x,y}}L, \,\, 
\hat{h} = \frac{h L}{\theta_c}, \,\, 
\hat{D}_x = \frac{D_x}{L\theta_c}, \,\, 
\hat{K} = K L^2
\end{equation}
where $s$ is the path length measured from the cell entrance, 
$D_x$ is horizontal dispersion, $\beta_{x,y}$ are horizontal and vertical beta functions, 
$h=\frac{1}{\rho}$ is the curvature function, $\rho$ is the bending radius, 
$K=\frac{1}{B\rho}\frac{\partial B_y}{\partial x}$ is the focusing gradient, 
$B\rho$ is the magnetic rigidity of the electron beam,  
and $\theta_c$ is the total bending angle of the cell. 
All $\hat{\cdot}$ quantities are dimensionless. 
The differential equations for the scaled linear lattice functions, 
$\hat{\beta}_{x,y}$ and $\hat{\eta}$, are
\begin{eqnarray} \label{eq:er4}
\frac{1}{2}\hat{\beta}_{x,y}^{''} + \hat{K}_{x,y}(\hat{s})\hat{\beta}_{x,y} - 
\frac{1+\hat{\alpha}_{x,y}^2}{\hat{\beta}_{x,y}} &=0, \\
\hat{D}_x^{''} + \hat{K}_x(\hat{s})\hat{D}_x &= \hat{h}(\hat{s}),  \label{eq:er4a}
\end{eqnarray}
where $\hat{\alpha}_{x,y}=-\frac{\hat{\beta}'_{x,y}}2$, 
$\hat{K}_x=\hat{K}+\hat{h}^2\theta_c^2\approx \hat{K}$,  $\hat{K}_y=-\hat{K}$, and 
$'$ and $''$ refer to first and second order derivatives with respect to $\hat{s}$, respectively. 
The approximation $\hat{K}_x\approx \hat{K}$ is valid when the deflection angle is small. 

The total deflection angle of the normalized cell 
is a constraint of the lattice cell optimization, which corresponds to the condition 
\begin{equation}\label{eq:er5}
\int_0^1 \hat{h}(\hat{s}) \mathrm{d} \hat{s} = 1. 
\end{equation}
The maximum quadrupole gradient is another important 
parameter that impacts the lattice cell design. High gradient would enable packing more cells over a given 
circumference and in turn reduces the emittance. In the optimization it is necessary to specify the upper limit of the 
quadrupole gradient. This corresponds to another constraint~\cite{HuangIPAC17}
\begin{equation}\label{eq:er6}
|\hat{K}(\hat{s})|\leq\hat{K}_\text{max} \equiv K_\text{max}L^2. 
\end{equation} 

Dimensionless radiation integrals can be defined using the 
dimensionless lattice functions 
\begin{subequations} \label{eq:er7}
\begin{align}
\hat{I}_2 &= \int_0^1 \hat{h}^2(\hat{s}) \mathrm{d} \hat{s} \\
\hat{I}_3 &= \int_0^1 |\hat{h}|^3(\hat{s}) \mathrm{d} \hat{s} \\
\hat{I}_4 &= \int_0^1 \hat{D}_x(\hat{s}) \hat{h}(\hat{s})(\hat{h}^2(\hat{s})+2\hat{K}(\hat{s})) \mathrm{d} \hat{s} \\
\hat{I}_5 &= \int_0^1 \hat{\mathcal{H}}(\hat{s}) |\hat{h}(\hat{s})|^3 \mathrm{d} \hat{s}
\end{align}
\end{subequations}
where $\hat{\mathcal{H}}=\hat{\beta}_x\hat{D}_x'^2+2\hat{\alpha}_x\hat{D}\hat{D}_x'+\hat{\gamma}_x\hat{D}_x^2$ 
is the normalized dispersion invariant, $\hat{\alpha}_x=\alpha_x$ and $\hat{\gamma}_x=\gamma_xL$ are normalized 
Courant-Snyder parameters. 
The equilibrium horizontal emittance and the rms energy spread of an electron storage ring are given by 
\begin{eqnarray} \label{eq:er1}
\epsilon &=& C_q\gamma^2 \theta_c^3 \frac{\hat{I}_5}{J_x\hat{I}_2}, \\
\sigma_\delta^2 &=& C_q\gamma^2 \frac{\theta_c}{L} \frac{\hat{I}_3}{2\hat{I}_2+\hat{I}_4}
\end{eqnarray} 
where $C_q\approx3.83\times10^{-13}$~m, $J_x=1-\frac{\hat{I}_4}{\hat{I}_2}$, and $\gamma$ is the relative energy factor. 

This problem is then to find the suitable 
functions $\hat{K}(\hat{s})$ and $\hat{h}(\hat{s})$ which satisfy Eqs.~\eqref{eq:er4}-\eqref{eq:er6} 
and minimize the emittance as given by
Eq.~\eqref{eq:er1} for any given set of deflection angle $\theta_c$ and $\hat{K}_\text{max}$.

\subsection{Optimization of the lattice cell}
A major goal of lattice cell optimization for storage ring light source is to achieve high photon 
beam brightness, $\mathscr{B}$, which is inversely proportional to the phase space volume of the photon beam~\cite{KIM198671}, 
\begin{equation} \label{eq:er10}
\mathscr{B}\propto\frac{1}{\sigma_{ph,x}\sigma_{ph,x'}\sigma_{ph,y}\sigma_{ph,y'}}, 
\end{equation}
where $\sigma_{ph,x}$, $\sigma_{ph,y}$, $\sigma_{ph,x'}$ and $\sigma_{ph,y'}$ are photon beam width or divergence in 
both transverse directions. They are related to electron beam dimensions in the phase space 
and the radiation distribution of a single electron through
\begin{subequations} \label{eq:er9}
\begin{align}
\sigma^2_{ph,xy} &= \sigma^2_{e,xy}+\sigma^2_{\lambda} \\
\sigma^2_{ph,x'y'}&= \sigma^2_{e,x'y'}+\sigma^2_{\lambda'}.
\end{align}
\end{subequations}
where the $\sigma_{\lambda}$ and $\sigma'_{\lambda}$ are the size and divergence of photon beam at the photon beam 
source point. 
The electron beam sizes, $\sigma^2_{e,xy}$ and $\sigma^2_{e,x'y'}$,
are related to the beam emittances, energy spread, and the 
lattice functions, 
\begin{subequations} \label{eq:er8}
\begin{align}
\sigma_{e,x} &=  \sqrt{\epsilon_{x}\beta_{x}+(D_x\sigma_\varepsilon)^2},  
& \sigma_{e,y} &=  \sqrt{\epsilon_{y}\beta_{y}},\\
\sigma_{e,x'}& =\sqrt{\epsilon_{x}\gamma_{x}+(D_x'\sigma_\varepsilon)^2}, 
& \sigma_{e,y'} &=  \sqrt{\epsilon_{y}\gamma_{y}}. 
\end{align}
\end{subequations}

In the cell optimization studies, we first used the brightness as the optimization objective. This 
would automatically include the impact of emittance and beta functions. 
However, the actual lattice cell design often uses quadrupole doublet or triplet at the end of the 
cell to tailor the beta functions at the location of insertion devices. 
Therefore, the emittance of the cell is not necessarily tied to the beta functions. 
For this reason, we also did lattice cell optimization to minimize the natural emittance only. 

\section{Optimization setup \label{secOptimSetup}}
As indicated in Eqs.~(\ref{eq:er4}-\ref{eq:er4a}), the linear lattice functions of a periodic cell are 
completely determined by the focusing function $\hat{K}(\hat{s})$ and the curvature function $\hat{h}(\hat{s})$. 
With the constraints on the two functions as given by Eqs.~(\ref{eq:er5}-\ref{eq:er6}), we can 
optimize the lattice cell in a very general manner. 
The objective of the optimization can be the brightness of the photon beam, or the equilibrium 
emittance of the electron beam.  
The variables for the optimization are parameters that specify the two functions over the length 
of the cell. 
In the following we will describe the optimization setup in more details.  

\subsection{Optimization Parameters and Objectives}
Although theoretically the focusing and curvature functions are smooth functions, they can be 
approximated with function values at a finite number of points separated by equal intervals. 
One way to represent the functions would be to specify the function values at the given control 
points. However, if the function values at these points are independently changed, 
the constraint in Eq.~(\ref{eq:er5}) will most likely be  violated, which requires additional 
adjustment to the function values, for example, by scaling, to satisfy the constraint. 
Another disadvantage of independently specifying function values at various points is that 
it is more difficult for the optimization algorithms to develop global patterns, such as symmetric 
patterns in the functions. 

\begin{figure} [ht]
\centering
\includegraphics[width=\linewidth]{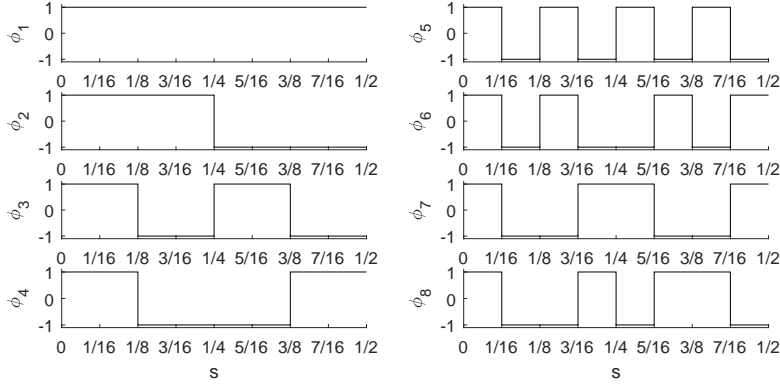}
\caption{The dipole field basis functions $\phi_n(\hat{s})$ for $m=3$ over the half cell in [0, $\frac12$]. The other half is in 
mirror symmetry about the $\hat{s}=\frac12$ point. }
\label{figure:basis_function}
\end{figure}

We assume the focusing function and the curvature function are both symmetric about the 
center of the cell. The full extent of the cell covers the range $\hat{s}\in [0, 1]$. Constrained by the Eq.~(\ref{eq:er5}), 
the curvature function $\hat{h}(\hat{s})$ with arbitrary distribution, over the half cell of $\hat{s}\in [0, 0.5]$, can be represented  
through a series of basis functions
\begin{eqnarray}\label{eq:hfield}
\hat{h}(\hat{s}) = \sum_{n=1}^N a_n \phi_n(\hat{s}),
\end{eqnarray}
where $N=2^m$, $m$ is an integer, %$n=1,2,\cdots,N$, 
and $\phi_n(\hat{s})$ are functions of piece-wise constant values of $1$ 
and $-1$. The basis functions are orthogonal to each other, i.e.,
\begin{eqnarray}
\int_0^1 \phi_i(\hat{s})\phi_j(\hat{s}) d\hat{s} = 0, \qquad \text{if}\quad i\neq j.
\end{eqnarray}
The basis functions $\phi_n$ for  $m=3$ is illustrated in Figure~\ref{figure:basis_function} as an example.
Except for $\phi_1(\hat{s})=1$ (uniform distribution), all basis functions integrate to zero over the zone 
of [0, 1]. 
Therefore $a_1$ is the only parameter that changes the total deflecting angle. 
The final shape of the function $\hat{h}(\hat{s})$ is determined by the coefficients of the 
other basis functions.

The focusing function $\hat{K}(\hat{s})$ is also represented by piece-wise constants in the same $N$ slices, 
but with different basis functions,
\begin{eqnarray}
\hat{K}(\hat{s}) = \sum_{n=1}^N b_n \psi_n(\hat{s}),
\end{eqnarray}
where $\psi_n(\hat{s})$ are 
\begin{equation}
\psi_i(\hat{s}) =
  \begin{cases}
    1       & \hat{s}\in(\frac{i-1}{2N},\frac{i}{2N}) \\
    0       & \hat{s}\not\in(\frac{i-1}{2N},\frac{i}{2N})
  \end{cases},
  i=1,2,3...N
\end{equation}
and the control parameters, $b_n$, are chosen with the constraint of not violating Eq.~(\ref{eq:er6}) in any slice. 
For a cell with $N=2^m$ slices with mirror symmetry about the cell center, there are $N-1$ control parameters 
for determining the dipole field $\hat{h}(\hat{s})$ and the quadrupole field $\hat{K}(\hat{s})$.

Given the curvature function $\hat{h}(\hat{s})$ and focusing function $\hat{K}(\hat{s})$, the natural emittance $\epsilon_n$ can be evaluated by Eqs.~(\ref{eq:er7}-\ref{eq:er8}). For the calculation of photon brightness $\mathscr{B}$, the vertical emittance $\epsilon_y$ due to the coupled motion from the horizontal plane should be considered. The horizontal emittance $\epsilon_x$ and vertical emittance $\epsilon_y$ can be expressed as
\begin{align}\label{eq:both emittance}
\epsilon_x &= \frac{1}{1+\kappa}\epsilon_n, \qquad
\epsilon_y = \frac{\kappa}{1+\kappa}\epsilon_n,
\end{align}
where $\kappa$ is the coupling ratio between the vertical and horizontal planes. Electron beam beam energy of 2 GeV and 
photon energy of 10 keV are assumed in the brightness calculation using Eq.~(\ref{eq:er10}-\ref{eq:er8}).

\subsection{The Particle Swarm Optimization Algorithm}
Optimization algorithms capable of finding the global minimum in a large parameter space is needed for the 
lattice cell optimization problem described in the above. 
In such cases, stochastic algorithms could be used. 
Stochastic optimization algorithms such as genetic algorithms~\cite{Bazarov2005, BorlandGA, YangGANSLS2}, and 
particle swarm optimization (PSO)~\cite{BorlandPSO, BaiPSO2011,Pang2014124, xiahuang2014} have been used on various accelerator problems. 
In this study, we use the PSO algorithm found in Refs.~\cite{Pang2014124, xiahuang2014} as it was demonstrated 
to have fast convergence due to the high diversity in the evaluated new solutions. 

In the PSO algorithm, each solution is considered as a moving particle in the control parameter space. 
A population of such particles are manipulated by the algorithm for many iterations. During each iteration, the 
position of each particle is updated by adding an amount called its velocity,
\begin{equation} \label{eq:er18}
\vec{x}_i^{t+1} = \vec{x}_i^t + \vec{v}_i^{t+1}
\end{equation}
where $\vec{x}_i^t$ and $\vec{v}_i^t$ are vectors that represent the position and the velocity of the $i^{th}$ particle 
at iteration $t$, respectively. The velocity is calculated as the weighted sum of three terms,
\begin{equation} \label{eq:er19}
\vec{v}_i^{t+1} = \omega\vec{v}_i^t + c_1(\vec{p}_i^t-\vec{x}_i^t) + c_2(\vec{g}^t-\vec{x}_i^t)
\end{equation}
where the three terms on the right-hand side, from left to right, represent the previous velocity, the 
distance between the present position and the position of the best solution in the history of this 
particle(i.e. personal best, $\vec{p}_i^t$), and the distance between the present position and the global 
best solution $\vec{g}^t$, respectively. 
Parameters $\omega$ and $c_{1,2}$  control the behavior of the algorithm and are given as in Ref.~\cite{xiahuang2014}. 
Mutation operation is also performed to a small fraction of randomly selected solutions.
After the initial position and velocity distributions are given, the particles move in the parameters space along trajectories according 
to the function evaluations and Eqs.~(\ref{eq:er18}-\ref{eq:er19}). 

\begin{figure} [ht]
\centering
\includegraphics[width=0.8\linewidth]{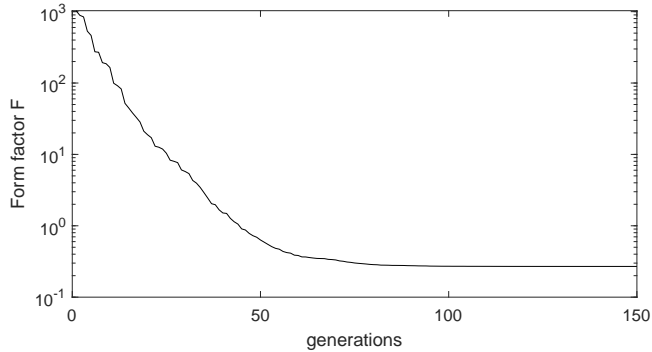}
\caption{The evolution of the best (lowest) form factor in a cell emittance minimization run by the PSO algorithm is shown 
as an indication of the convergence performance of the algorithm. The setup includes 63 control parameters for each solution 
and 1500 solutions in the population. }
\label{figure:conv}
\end{figure}

The PSO algorithm applied to the lattice cell optimization problem has good convergence performance as 
indicated by the example shown in Figure~\ref{figure:conv}, which shows the evolution of the best(lowest) 
form factor, $\mathcal{F}$, over 700 iterations. 
There are $63$ control parameters in this setup and 1500 solutions in the population.
%% a figure here

\section{Optimization results \label{secResults}}
\subsection{Brightness $\mathscr{B}$ as the optimization objective}
When photon brightness, $\mathscr{B}$, is used as the optimization objective, the electron beam energy and photon energy need to 
be specified. In this study we assumed a 2 GeV electron beam and a 10 keV photon energy. The undulator length is 
assumed $L=2.5$~m in the calculation of single photon divergence.

\subsubsection{The impact of coupling ratio}
With the beam on or near the linear difference coupling resonance, the natural emittance is split between the horizontal 
and vertical planes through Eqs.~(\ref{eq:both emittance}). 
Therefore, the coupling ratio $\kappa$ could be a key parameter that significantly impacts the brightness optimization. 
We investigated the effect of the coupling ratio by performing the cell optimization for maximum brightness with 
the coupling ratio changed over a large range. The results are summarized in  Table~\ref{table:different_coupling}, which lists 
the natural emittance, the horizontal partition number, and the horizontal phase advance for the optimized cell for various 
coupling ratios. 
Surprisingly, the resulting optimized cells are almost identical when the coupling ratio is changed. The only exception is when 
$\kappa=0$, in which case the vertical beam emittance is zero and hence the vertical beta function is not involved in the 
brightness calculation. 
For all cases with $\kappa>0$, the optimized cell has approximately  $\Phi_x=163^\circ$ for the 
horizontal phase advance and $\Phi_y=105^\circ$ for the  
vertical phase advance. The optimized horizontal partition number is $J_x=1.28$. 
As will be shown in the next section, these optimized parameters strongly depend on the maximum quadrupole gradient. 
\begin{table} [ht]
\centering
\caption{The optimized parameters for different coupling ratios in the brightness optimization. 
 The maximum normalized quadrupole gradient is $\hat{K}_\text{max}=13$. 
 The emittance value is for a 2~GeV beam, with cell length of $L=1.0$~m and deflection angle of $\theta_c=1^\circ$.  
 The form factor can be calculated with $\mathcal{F}=\frac{J_x \epsilon_n}{0.671\,\text{[pm]}}$.}
\begin{tabular}{c|c|c|c}
    \hline
    Coupling & Emittance & Horizontal partition & Phase advance \\
    $\kappa$ & $\epsilon_n$ (pm) & $J_x$ &  $\Phi_x$/$\Phi_y$ ($^{\circ})$\\ \hline
    0 & 0.156 & 1.17 & 136/97  \\ \hline
    0.1 & 0.206 & 1.28 & 163/105 \\ \hline
%    0.2 & 0.203 & 1.27 & 163/103 \\   \hline
%    0.3 & 0.198 & 1.27 & 163/103 \\ \hline
%    0.4 & 0.201 & 1.27 & 163/104 \\ \hline
    0.5 & 0.206 & 1.28 & 163/105 \\ \hline
%    0.8 & 0.201 & 1.29 & 163/105 \\ \hline
    1.0 & 0.202 & 1.28 & 163/105 \\ \hline
    \end{tabular}
    \label{table:different_coupling}
\end{table}

Because  the brightness optimization is not sensitive to the coupling ratio, in the following the coupling ratio is set to 
a constant value of $\kappa = 1$.

\subsubsection{The impact of maximum quadrupole gradient}
As discussed earlier, higher quadrupole gradient allows placing more focused bending magnets in a ring; and this is 
what enabled the MBA lattice cell to substantially 
reduce the electron beam emittance from the traditional DBA-based design approach. 
It is expected that the brightness should be strongly dependent on the quadrupole field distribution, $\hat{K}(s)$. 
But how would the maximum quadrupole gradient affect the quadrupole field distribution is not clear before the numerical 
optimization is conducted.

In this section, the effect of the maximum quadrupole gradient on the brightness optimization is investigated. 
In the optimization study, the cell length is set to $L=1$~m and the cell deflection angle is $\theta=1^{\circ}$. 
The slice number for the cell is $N=64$. 
Because of the mirror symmetry about the cell center, there are $31$ control parameters for dipole field distribution, 
$\hat{h}$, and $32$ control parameters for quadrupole field distribution, $\hat{K}$.
The normalized maximum quadrupole gradient, $\hat{K}_\text{max}$, is varied from 13 to 208. 

\begin{figure} [htp]
\centering
\includegraphics[width=\linewidth]{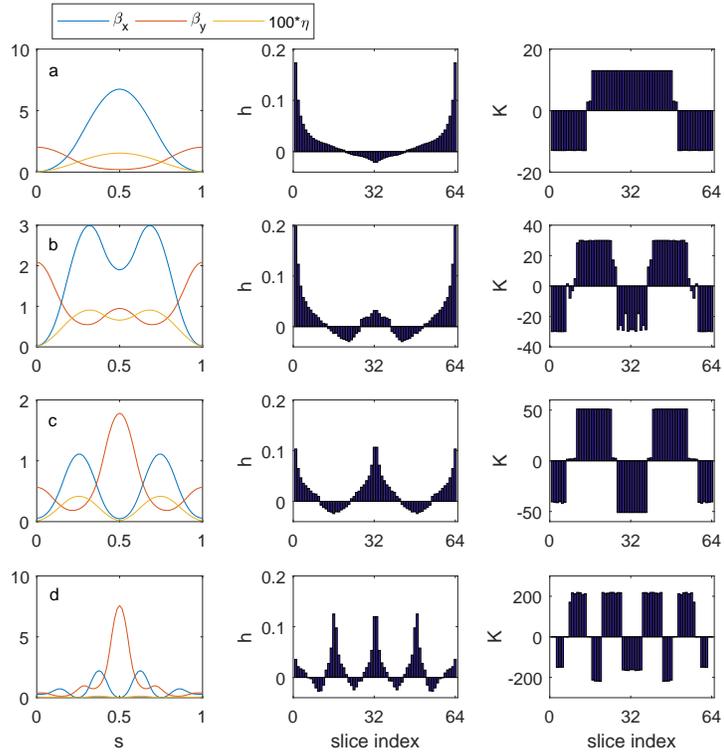}
\caption{The optimization results in terms of the optics functions (left column), 
dipole field distribution (middle column), and quadrupole field distribution (right column) for 
brightness optimization with different maximum quadrupole gradients. The four rows 
represent four levels of $\hat{K}_\text{max}$: 
(a) $\hat{K}_\text{max}=13$; (b) $\hat{K}_\text{max}=30$; (c) $\hat{K}_\text{max}=52$; (d) $\hat{K}_\text{max}=208$. 
}
\label{figure:Kmax}
\end{figure}

Figure~\ref{figure:Kmax} shows the optics functions of the optimized cell (left column), the corresponding 
dipole field distribution,  $\hat{h}(s)$ (center column), and the quadrupole field distribution, $\hat{K}(s)$ (right column) 
for four levels of $\hat{K}_\text{max}$. 
The corresponding emittance, horizontal partition number, and horizontal phase advances parameters are listed in 
Table~\ref{table:different_Kmax}. 

The case in Figure~\ref{figure:Kmax}(a) ($\hat{K}_\text{max} = 13$) resembles the general case of one focused dipole cell. 
The quadrupole field in the optimized cell naturally groups into one focusing magnet and one defocusing magnet (if we consider 
the cell ends as the center of the defocusing magnet). The gradients of both magnets are at the maximum value. 
There is one minimum and one maximum in the beta function of each transverse plane. Because the brightness, which is the 
objective function to be minimized, is 
calculated at the ends of the cell, the minimum of horizontal beta function is at the ends. 
The dipole field function develops an interesting distribution over the cell length, which includes two important features: 
the longitudinal dipole gradient and the anti-bends. 
The maximum bending field is at the cell ends, where the normalized curvature is $\hat{h}\approx 10$, which corresponds to a dipole 
field of 1.17~T for a cell bending angle of $1^\circ$ on a 1-m long cell and a 2~GeV beam.
The negative bending angle is $-21.3\%$ of the total cell bending angle. 

As $\hat{K}_\text{max}$ is increased to 30, the dipole and quadrupole field distributions on the cell have developed 
features of two separate cells. When it is increased to 52, which is 4 times of the case in Figure~\ref{figure:Kmax}(a), 
the field distributions are very similar to two cells scaled from case Figure~\ref{figure:Kmax}(a). 
The vertical beta function does not look like two identical cells because the vertical beta function affects 
the optimization objective function only at the ends and is thus not sufficiently constrained. 
When $\hat{K}_\text{max}$ is increased by another factor of 4, to 208, the field distributions appear to 
consist of four cells of the type found in case Figure~\ref{figure:Kmax}(a). The horizontal beta and dispersion function do not 
show ideal periodicity, probably because the number of slices is not enough for it to exhibit the full features of the 
case (a) cell type. 

The optimized field distributions as shown in Figure~\ref{figure:Kmax} when the maximum quadrupole gradient is varied 
clearly indicate that the cell type found in Figure~\ref{figure:Kmax} (a) is a fundamental building block of 
low emittance lattice aimed at high brightness. 
When the maximum quadrupole gradient is high enough, the natural tendency for obtaining high brightness is 
to split the cell into more cells of this type. 
%% a figure here
%% a table here
\begin{table} [ht]
\centering
\caption{Key parameters of the optimized cell in the brightness optimization 
for different $\hat{K}_\text{max}$. The coupling ratio is set to $\kappa=1$. 
 The emittance value is for a 2~GeV beam, with cell length of $L=1.0$~m and deflection angle of $\theta_c=1^\circ$. 
 The form factor can be calculated with $\mathcal{F}=\frac{J_x \epsilon_n}{0.671\,\text{[pm]}}$.}
\label{table:different_Kmax}
\begin{tabular}{ c | c | c | c }
    \hline
     & Emittance/pm & Horizontal partition & phase advance \\
    $\hat{K}_\text{max}$ & $\epsilon_n$ & $J_x$ &$\Phi_x$/$\Phi_y$ ($^{\circ})$ \\ \hline
    13  & 0.207 & 1.28 & 163/107  \\ \hline
    30  & 0.208 & 1.56 & 167/68 \\ \hline
    52  & 0.042 & 1.40 & 307/168 \\  \hline
    208 & 0.013 & 1.50 & 541/370 \\ \hline    
    \end{tabular}
\label{table:different_Kmax}
\end{table}

The quadrupole field distribution in Figure~\ref{figure:Kmax} shows that a smooth function for the 
gradient does not have any advantage in increasing the brightness. The $\hat{K}(s)$ function tends to group 
in areas with maximum focusing or defocusing strengths. Therefore, in the following we model the gradient 
function as discrete quadrupole magnets, varying only their gradient values and locations. 

\subsubsection{Transforming one cell into two cells}
The splitting of a lattice cell into two basic cell types as we found in the previous section is an 
important and interesting phenomenon. Because distributed quadrupole fields over the entire cell is not 
realistic and does not have any advantage, we decided to further investigate the phenomenon with a model of 
discrete quadrupoles. 

\begin{figure} [htp]
\centering
\includegraphics[width=0.8\linewidth]{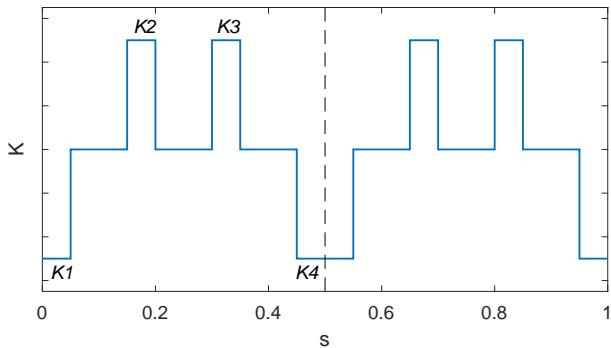}
\caption{The quadrupole configuration for the study of cell splitting with discrete quadrupoles, 
$K1$, $K2$, $K3$, and $K4$, which are arranged about the cell center with mirror symmetry. }. 
\label{figure:K_arrangement}
\end{figure}

In this model eight quadrupoles are placed in the cell in a symmetric configuration about the cell center, 
as shown in Figure~\ref{figure:K_arrangement}. 
All quadrupole lengths are set to 0.05. The position of quadrupoles $K1$ and $K4$ are fixed at the cell end, or 
cell center, respectively, while $K2$ and $K3$ can be freely moved, up to the edges of $K1$ or $K4$. $K2$ and $K3$ cannot overlap or 
cross each other. 
The dipole field distribution is represented in the same manner as in the previous section. 

\begin{figure} [htp]
\centering
\includegraphics[width=\linewidth]{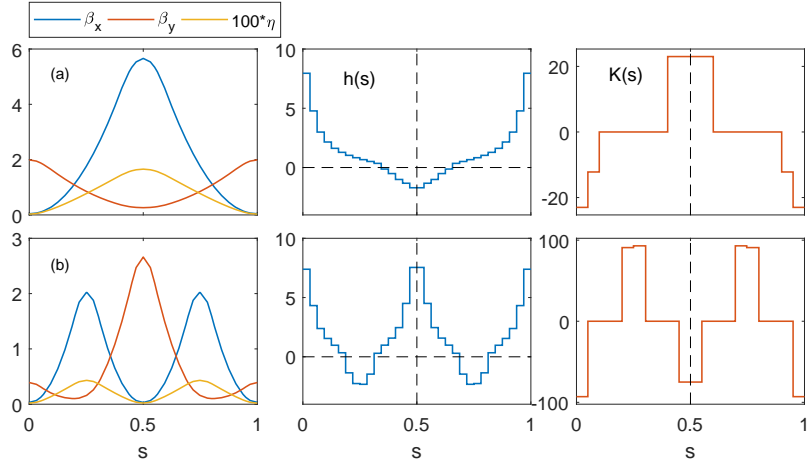}
\caption{The optics function (left column), dipole field profile (center column), and quadrupole distribution (right column) 
for the optimized cell with: (a) $\hat{K}_\text{max}=23$; (b) $\hat{K}_\text{max}=93$. 
The length is 0.05 for all quadrupoles. }
\label{figure:Kmax_23_93}
\end{figure}

The brightness optimization is performed for a series of cases when the maximum quadrupole gradient is varied from 
$\hat{K}_{max}=23$ to $\hat{K}_{max}=93$. 
The optics functions and field distribution functions for the cases of $\hat{K}_{max}=23$ to $\hat{K}_{max}=93$ are shown 
in Figure~\ref{figure:Kmax_23_93}. 
For the case with $\hat{K}_{max}=23$, quadrupole $K2$ moves to the cell end to join $K1$, and quadrupole $K3$ moves 
to the cell center to join $K4$. The dipole field distribution is similar to case Figure~\ref{figure:Kmax} (a) in 
the previous section, consisting one basic cell. The horizontal phase advance is $\Phi_x=164.0^\circ$. 
For the case with $\hat{K}_{max}=93$, however, quadrupole $K2$ and $K3$ move toward each other and meet at 
the $\hat{s}=0.25$ point. In this case, the cell is split into two basic cells that are similar to the cell type 
as in the case with $\hat{K}_{max}=23$. 
The horizontal phase advance becomes $\Phi_x=309.0^\circ$.

\begin{figure} [htp]
\centering
\includegraphics[width=\linewidth]{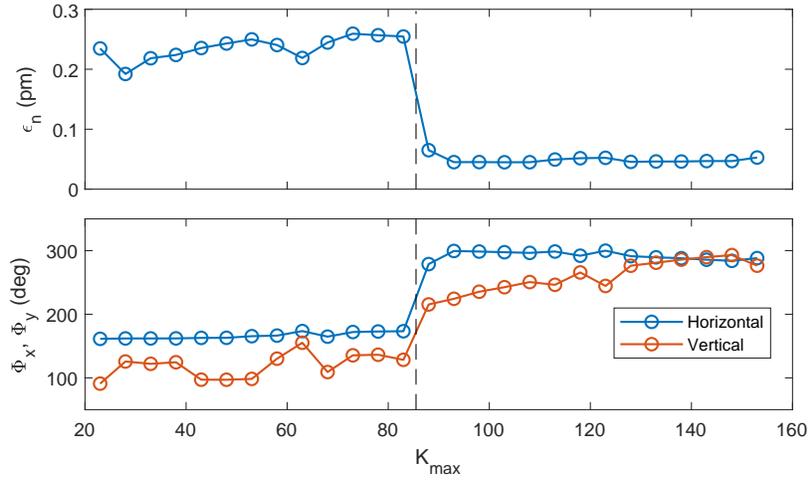}
\caption{ Natural emittance (top), for a 2~GeV beam, with cell length of $L=1.0$~m and deflection angle of $\theta_c=1^\circ$,
and phase advances (bottom) for optimized cells with various $\hat{K}_\text{max}$. }
\label{figure:Kmax_varation}
\end{figure}

The natural emittance, corresponding to a one-meter cell with bending angle $\theta_c=1^\circ$ and beam 
energy of 2~GeV, and the phase advances as functions of the maximum gradient are shown in 
Figure~\ref{figure:Kmax_varation}. Interestingly, before the maximum gradient is 
high enough for the cell to split, the phase advance and the natural emittance vary only slightly, 
despite significant changes to $\hat{K}_\text{max}$. 
This again indicates that the basic cell type as seen in Figure~\ref{figure:Kmax_23_93} (a) is an 
efficient fundamental building block of low emittance lattice cells.

%% a figure here

%% a figure here

\subsection{Using natural emittance as the optimization objective \label{objection_emittance}}

In the practical lattice design of MBA cells for diffraction limited rings, the optics of 
center bending magnets are connected to the insertion device straight section through a matching section. 
The beta functions at the insertion devices are thus decoupled from the 
the optics of the center bends, which dominate in the determination of the cell emittance. 
Therefore, the emittance minimization 
with the optics around the bends can be done separately from the photon brightness optimization.  

The optics around the center bends may consist of periodic sub-cells. Each sub-cell 
(which we will call a cell in the rest of this section) is a  alternating-gradient 
structure with stable optics. We would like to optimize the dipole field distribution and the linear topics on the cell 
to achieve the minimum emittance. 
In the cell optimization setup the dipole field distribution is represented by the same scheme as used for 
the brightness optimization (see Eq.~\eqref{eq:hfield}). Two thin quadrupoles are used for the alternating 
gradient focusing scheme, one is at the cell center, the other is split into two and placed at both 
ends. The integrated gradients of the two thin quads are used as control parameters. 
Essentially, we are optimizing the optics and the distributed dipole field on a FODO cell. 

This part of the study is similar to the numerical optimization of longitudinal-gradient bends found in 
Ref.~\cite{STREUN201598}, but with a few important differences. 
First, in Ref.~\cite{STREUN201598}, the beta function and dispersion function values at the center of the 
bend are used as control parameters. Matching quadrupoles are required to realize the optimized beta 
and dispersion functions. In this study, the quadrupole strengths are used as control parameters and 
hence the optics on the cell is completely determined. 
Second, in this cell the dipole cell is distributed throughout the cell, not only inside the dipole 
magnet. 
Third, Ref.~\cite{STREUN201598} only considered positive bending field, while in this study the dipole 
field can be negative. 
Although the benefits of negative bends have been understood in previous studies~\cite{STREUN2014148}, 
this study could reveal features of the optimal dipole field distribution with both longitudinal dipole gradient 
and anti-bends.

\subsubsection{Effect of dipole field slice number}
As pointed out in Ref.~\cite{STREUN201598}, when using the longitudinal dipole field profile to minimize the 
emittance, the dipole field strength tends to diverge to infinity at the center of the dipole. 
Therefore, it is necessary to impose a limit on the maximum dipole field strength for the optimization 
results to be practical. This can done by setting the maximum dipole field strength directly to a value 
deemed reasonable. %, e.g., 5~T. 
In this study, we took a different approach to limit the dipole field strength. 
Instead of setting the maximum dipole field strength, we set the width of the slice with the maximum 
dipole field to a finite value. 
This is done naturally in our optimization setup as we 
slice the cell into $N$ equal pieces with constant field on each slice. 
The approach of using a finite number of slices with piece-wise constant field is reasonable in 
practical terms as in reality the cell would consist of a series of dipole magnets with constant 
field strengths and finite widths.   
This approach is also convenient as it allows us to use the normalized parameters 
defined in Section~\ref{refNormCell} without the need to convert an absolute dipole field strength to 
the normalized parameter under specific assumptions of beam energy, cell length, and cell bending angle. 
In this sub-section we investigate the dependence of cell emittance optimization on the number of 
slices of the dipole field. 

For the normalized FODO cell to be stable, the maximum integrated quadrupole gradient is 
$\hat{K}\Delta \hat{L}=4$, when both the horizontal and vertical phase advances on the cell are 
$180^\circ$ (in the case when the focal lengths of the QF and QD magnets are equal). 
In the study of the effect of the number of slices, we set the limit of the integrated 
quadrupole gradient of the QF and QD magnets to $\hat{K}_\text{max}\Delta \hat{L}=4$. 

\begin{figure} [htp]
\centering
\includegraphics[width=\linewidth]{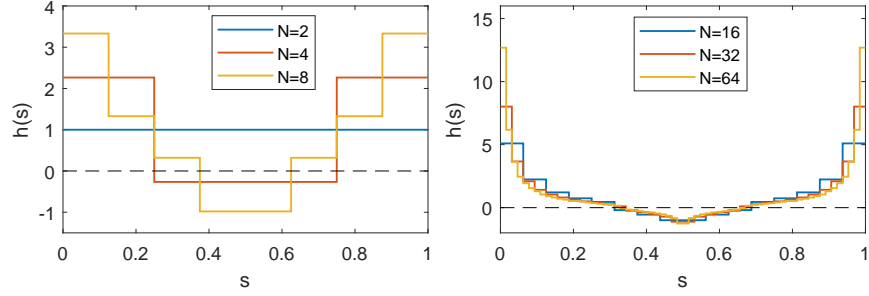}
\caption{The optimized dipole field profiles, $\hat{h}(s)$, 
with various slice numbers in the cell in 
the emittance optimization. }
\label{figure:comparing_different_slice}
\end{figure}

\begin{figure} [htp]
\centering

\includegraphics[width=\linewidth]{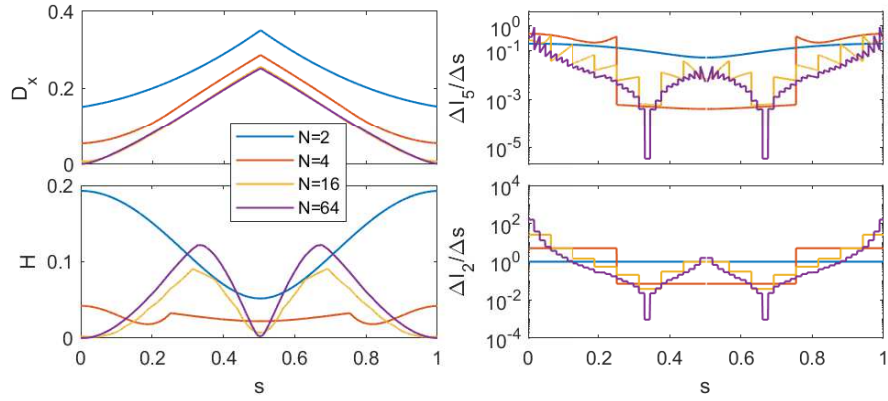}
\caption{The distribution of dispersion-related parameters in the optimized cell  
for $N=2$, 4, 16, 64. Top left: dispersion function, $D_x$; 
top right: contribution to the $\hat{I}_5$ integral per unit length, $\frac{d\hat{I}_5}{d\hat{s}}$; 
bottom left: dispersion invariant; %$\mathcal{H}=\frac1\beta [D_x^2+(\alpha_x D+\beta D')^2]$; 
bottom right: contribution to the $\hat{I}_2$ integral 
per unit length, $\frac{d\hat{I}_2}{d\hat{s}}$.
}
\label{figure:comparing_different_slice2}
\end{figure}

The number of slices was set to $N=2^m$, with $m=1, 2, \cdots, 6$. For each case, there are 
$2^{m-1}-1$ control parameters for the dipole field profile and 2 control parameters for the quadrupoles. 
The optimized dipole field profiles for all cases of slice numbers are shown in 
Figure~\ref{figure:comparing_different_slice}. 
The corresponding emittance form factors and some other related parameters are shown in 
Table~\ref{table:different_slice}.  
Figure~\ref{figure:comparing_different_slice2} shows the dispersion function, the 
dispersion invariant $\hat{\mathcal{H}}$,  %$\mathcal{H}=\frac1\beta [D_x^2+(\alpha_x D+\beta D')^2]$, 
and the  contributions to the $\hat{I}_2$ and $\hat{I}_5$ integrals, respectively, 
for the $N=2$, 4, 16, and 64 cases. 

For the case of $N=2$, the dipole field is constant over the cell length. 
The optimized cell has a horizontal phase advance of $\Phi_x=141^\circ$, which is 
very close to the theoretic minimum value of approximately $140^\circ$~\cite{SYLeeAPFODO}. 
The dipole field profiles for the $N\ge 4$ cases clearly reveals the diverging trend of the maximum dipole 
field strength when the cell is divided into more equal slices in the optimization setup. 
The emittance reduction is achieved by both reducing the 
$\hat{I}_5$ integral and increasing the $\hat{I}_2$ integral. 
The rate of emittance reduction with the increasing number of slices 
decreases. For example, from $N=8$ to $16$, the emittance is reduced by nearly 50\%, 
while from $N=32$ to 64, the reduction is only 33\%. 
In the mean time, the $\hat{I}_3$ integral, which affects the equilibrium momentum spread, 
increases at a steady pace with the doubling of slices. This indicates that 
increasing the slice number, or equivalently, 
allowing a higher maximum dipole field, has only limited applicability in 
lattice cell performance improvement, even if it is not limited by the technical difficulty 
in achieving the high bending field. 

It is noted that anti-bends are present in all cases with $N>2$. 
For $N\ge 4$ cases, the dispersion invariant develops a double-hump distribution which 
peaks at about $\hat{s}=\frac13$ and $\frac23$, where the curvature function $\hat{h}(s)$ crosses 
zero. 
With a large number of slices in the cell ($N\ge 8$), the phase advance tends to the high 
end, in order to minimize the emittance. 

Additional studies showed that if the maximum dipole field in the slices at the cell ends ($\hat{s}=0$ and 1) 
is given at a fixed value, the optimal dipole field distribution 
does not change when more  slices are introduced in the rest of the cell. This confirms that 
the variation of the optimized distribution with respect to the number of cells is dominated by 
the peak dipole field.

\begin{table} [h]
\centering
\caption{The optimized emittance form factor, horizontal phase advance, 
ratio of anti-bends, and radiation integrals
in the cell emittance optimization 
for various slice numbers. }
\begin{tabular}{c|c|c|c|c|c|c}
    \hline
    Slices  & Form fac. & phase adv. & anti-bend &  integral  & integral & integral  \\
     $N$ & $\mathcal{F}$ & $\Phi_x$~($^\circ$) & $\theta_-/\theta$ & $\hat{I}_5$ & $\hat{I}_2$ & $\hat{I}_3$   \\ \hline
    2 & 5.88 & 141 & 0 & 0.1266 & 1.00 & 1.00  \\ \hline
    4 & 3.00 & 156 & 0.133 & 0.1681 & 2.60 & 5.82   \\ \hline
    8 & 1.16 & 165 & 0.245 & 0.0872 & 3.48 & 10.07  \\   \hline
    16 & 0.59 & 171 & 0.218 & 0.0551 & 4.32 & 18.39  \\ \hline
    32 & 0.36 & 173 & 0.184 & 0.0433 & 5.56 & 36.27 \\ \hline
    64 & 0.24 & 176 & 0.159 & 0.0379 & 7.33 & 73.81 \\   \hline
    \end{tabular}
\label{table:different_slice}
\end{table}

\subsubsection{The effect of maximum quadrupole gradient on the optimization} 
In the previous subsection, the emittance optimization was done with the quadrupole gradient limited  
by FODO cell optics stability requirement. 
It could be useful to study how the optimal dipole field distribution depends on the maximum quadrupole 
gradient. 
The cell emittance optimization was repeated with various maximum integrated quadrupole gradients. 

\begin{figure} [htp]
\centering
\includegraphics[width=0.8\linewidth]{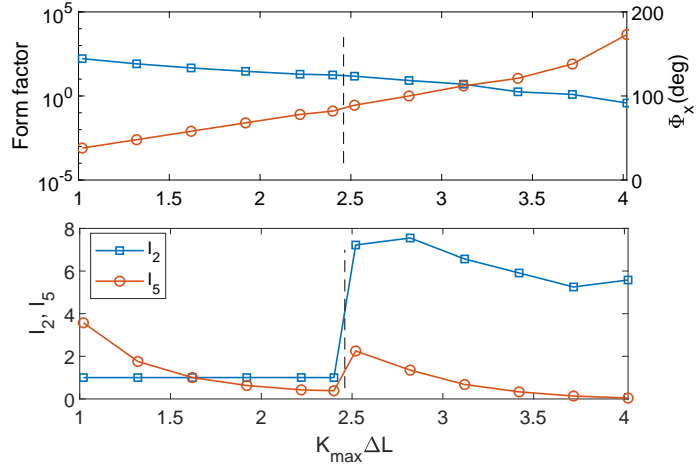}
\caption{Optimized cell parameters in the emittance optimization 
as the maximum integrated gradient $\hat{K}_\text{max}\Delta L$ is changed.
Top: the emittance form factor $\mathcal{F}$ (blue squares) and the horizontal phase advance (red circles); 
Bottom: radiation integral $I_2$ and $I_5$.  
The dashed vertical line 
indicates a sudden change of dipole field profile between $\hat{K}_\text{max}\Delta L=2.4$ and $2.55$. }
\label{figure:pars_Kmax}
\end{figure}

Figure~\ref{figure:pars_Kmax} shows the emittance form factor and the horizontal phase advance (top) of the 
optimized cell and the radiation integrals $I_2$ and $I_5$ (bottom) at various maximum integrated quadrupole strengths.  
As the maximum gradient is reduced, the phase advance on the cell decreases, while  
the minimum emittance achievable increases. 

\begin{figure} [htp]
\centering
\includegraphics[width=\linewidth]{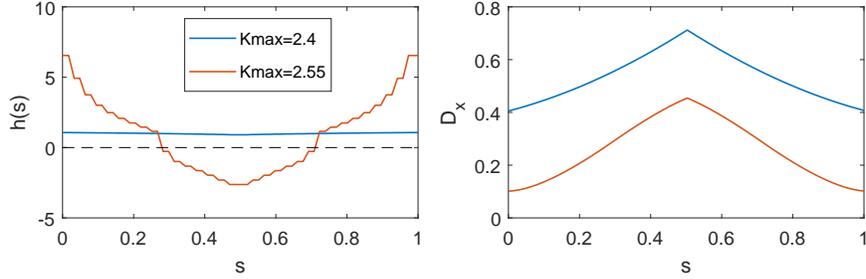}
\caption{The optimization results for before and after the dipole field distribution transitions from  
flat to with longitudinal gradient and anti-bend when $\hat{K}_\text{max}\Delta \hat{L}$ is increased from $2.4$ 
to $2.55$. 
Left: dipole field profile $\hat{h}(s)$; 
right: the horizontal dispresion function. }
\label{figure:kmax_imp}
\end{figure}

The optimized dipole field profile varies with the maximum quadrupole gradient. 
Interestingly, between the integrated gradient of $\hat{K}_\text{max}\Delta \hat{L}=2.4$ and $2.55$, the dipole 
field profile drastically changed. At $\hat{K}_\text{max}\Delta \hat{L}=2.4$ or below, the dipole field distribution is 
nearly flat, while at $2.55$ or above, there is large variation of dipole field on the cell and negative 
bending is present. 
The transition can be clearly seen in the $I_2$ and $I_5$ plot of Figure~\ref{figure:pars_Kmax}. 
However, the emittance form factors for the $\hat{K}_\text{max}\Delta \hat{L}=2.4$ and $2.55$ cases 
differ only slightly. 
The dipole field profiles  and the corresponding dispersion functions 
for the two cases are 
shown in Figure~\ref{figure:kmax_imp}. 

\section{Conclusion \label{sec:Conc}}
We have studied the optimal low emittance lattice cell structure with global numerical optimization of the general 
dipole and quadrupole field distributions over the cell. 
The normalized cell functions are used in the study, such that the optimized lattice cell is applicable to 
rings with different sizes and beam energies by scaling. 

Optimization is first done using the photon brightness of an insertion device as the objective while both 
the quadrupole and dipole field distributions are varied as control parameters. 
The maximum strength of the quadrupole gradient is limited as a main constraint in the optimization. 
It was found that with a relatively low quadrupole gradient limit, the linear optics of the optimized cell 
is similar to that of a FODO cell and 
the dipole field distribution  develops a pattern with longitudinal gradient and negative bending. 
When the maximum quadrupole gradient is increased sufficiently (by a factor of 4 of the low gradient case), 
the cell structure automatically split into 
two cells of the same type as the low quadrupole gradient case. 
When the maximum quadrupole gradient is further increased to a sufficient high level, the two cells 
further split into four cells. 
The cell splitting behavior indicates that the FODO cell type with dipole field variation is a fundamental 
cell type for low emittance ring lattices. 

We further studied the FODO cell optimization using the natural emittance as the objective. The integrated gradients of 
the 
focusing and defocusing quadrupoles and the dipole field distribution are the control parameters. 
The dipole field distribution is allowed to change over the entire cell, but with a limited number of  
constant-field slices. The finite number of slices naturally limits the maximum field strength. 
The optimized cell structure again shows longitudinal dipole field variation with negative bending. 

\section*{Acknowledgments}
  Work was supported by the U.S. Department of Energy, Office of
  Science, Office of Basic Energy Sciences, under Contract No.
  DE-AC02-76SF00515.

\section*{References}

\bibliography{mybibfile}

\end{document}